%% file: bullet.tex
\documentclass[final,3p,times,twocolumn]{elsarticle}

\usepackage{amssymb}
\usepackage{amsmath}

\usepackage{lineno}

\usepackage{array}
\usepackage{multirow}

\journal{Astroparticle Physics }

\newcommand \sK {\mathcal{K}}

\newcommand \bx {{\bf{x} }}

\newcommand \br {{\bf{r} }}

\begin{document}

\begin{frontmatter}

\linenumbers


\title{Many body gravity and the bullet cluster}
\author{S. Ganesh}
\ead{gans.phy@gmail.com}
\affiliation{organization={Independent Researcher},
           city={Bangalore},
           postcode={560024}, 
           country={India}}





\begin{abstract}
	Many body gravity (MBG) is an alternate theory of gravity, which has been able to explain the galaxy rotation curves, the radial acceleration relation (RAR) and the wide binary stars (WBS).
	The genesis of MBG is a novel theory, which models systems with thermal gradients, by recasting the variation in the temperature as a variation in the metric. Merging the above concept with Einstein's gravity, leads to the theory of thermal gravity in 5-D space-time-temperature. 
	Thermal gravity when generalized for partially thermalized systems, results in the theory of many body gravity. 
	The bullet cluster is supposed to be a smoking gun evidence for the presence of dark matter.
	However, this work demonstrates that the MBG theory can explain the weak gravitational lensing effect of the bullet cluster, without the need for yet undiscovered baryonic matter or dark matter.
\end{abstract}



\end{frontmatter}





\section{Introduction}
\label{sec:intro}
	Sir Issac Newton first proposed the laws of gravity, which were modified by Einstein about a hundred years ago in the form of the general theory of relativity. 
	The field equations of the general theory of relativity~\citep{Ein1} model gravity very accurately. Several experiments have been conducted~\citep{shap1, hol, sch, foma, shap2,ber}, which have validated Einstein's theory.
Gravitational lensing~\citep{lens1, lens2} and the recent detection of the gravitational waves~\citep{ligo1, ligo2} have further established the accuracy of Einstein's general theory of relativity.
	However, there are many phenomena, such as the galaxy rotation curves~\citep{rub1, rub2, galaxy1, galaxy2} and the gravitational lensing of the bullet cluster~\citep{bullet1, bullet2, bullet3, bullet4}, which have not yet been explained by Einstein's gravity, 

Galaxy rotation curves show that the velocity of stars in a galaxy is greater than expected when we go radially outward.
The question arises as to whether any undiscovered mass is responsible for the stars rotating faster.
This gives rise to the theory of dark matter~\citep{dark1, dark2, dark3, dark4, dark5}.
A second school of thought holds that there is no dark matter, but the laws of gravity must be adjusted~\citep{mond1, mond2, mond3, mond4, mond5, entropic, moffat1, moffat}.
	In Ref.~\cite{gans8}, a novel theory, namely the MBG theory, was proposed to explain the galaxy rotation curves.
	The MBG theory falls within the category of modified gravity. However, importantly, the theory has not been developed with the goal of explaining galaxy rotation curves or dark matter, unlike many other such theories. 
The existence of a fifth force is not assumed either.
The MBG theory originated from the pursuit of modeling the spatial and temporal variations of thermal systems.
It was postulated that the spatial and temporal thermal variations can be modeled by recasting the thermal variations as a variation in the metric in 5-D space time temperature~\citep{gans6,gans7}. 
Since thermal systems can be produced on earth, the concept behind the MBG theory can in principle be validated experimentally due to the much lower energy involved~\citep{gans6,gans7}. Moreover, the theory is manifestly relativistic.
	The MBG theory could also explain other phenomena such as the RAR, and the WBS systems~\citep{gans8}. 

The bullet cluster (IE0657) has been viewed as a smoking gun evidence for the existence of dark matter. The bulk baryonic matter and the bulk of the matter inferred via weak gravitational lensing are in different spatial regions~\citep{bullet1, bullet2, bullet3, bullet4}. 
The inter cluster gas (ICG), which forms the bulk of the baryonic matter, is located in-between the colliding galaxy clusters. 
However, the majority of the total mass, as determined by the weak gravitational lensing, lies coincident with the galaxy clusters. 
This makes it a challenge for modified gravity theories to explain, unless there is still undiscovered baryonic matter.  
However, it is now demonstrated that the theory of MBG is able to explain the mass inferred from gravitational lensing, i.e., the "dark matter", without requiring additional baryonic matter.
In this work, we show that a 3-D gas distribution around the bullet cluster, can account for the weak gravitational lensing phenomenon, within the framework of the MBG theory.
We demonstrate that the 3-D gas distribution leads to an effective mass distribution that is comparable to the mass estimated through weak gravitational lensing.
The effective mass is manifested due to the additional terms in the MBG field equations.

	The concept of recasting the variation in the temperature as a variation in the metric was first introduced in Ref.~\cite{gans5}, for modeling systems in local thermal equilibrium. The quark-antiquark potential in a Quark Gluon Plasma (QGP) with thermal variations, was calculated using the Anti De-Sitter/Conformal Field Theory (AdS-CFT) correspondence. The concept was placed on firmer grounds in Ref.~\cite{gans6}, using the Polyakov loop, the partition function and the geodesic equation. 
Calculations were performed in the field theoretic domain to calculate the pressure and the energy density of a scalar gas system, possessing spatial thermal gradients. 
In Ref.~\citep{gans6}, it was explained how to interpret the geodesic equation in the context of a curved space caused by thermal variations.
	Subsequently, the formulation was extended to thermal systems with temporal variations in a 5-D space-time-temperature framework~\citep{gans7}.
Since both thermal variations and gravity are modeled as a variation in the metric, it is natural to attempt to encapsulate both the phenomena in a common unified framework.
	This was carried out in Ref.~\citep{gans7}.
The encapsulation of the two phenomena in a unified framework, lead to novel phenomena like the spontaneous symmetry breaking of scalar fields under a very strong gravitational field~\citep{gans7}.
This is however subject to the scalar field being non-minimally coupled with the Ricci scalar~\citep{gans7}. 
The unified framework finally culminated in the MBG theory, which was able to explain the galaxy rotation curves, the RAR and the WBS systems~\citep{gans8}.

	The rest of the article is as follows. 
	An overview of the MBG theory developed in Ref.~\cite{gans8} is given in Sec.~\ref{sec:MBG}. The solution of the field equations in the MBG theory, the calculation of the effective mass that bends the path of light, and other formal developments are also covered in Sec.~\ref{sec:MBG}. An error in the solution of one of the field equations in Ref.~\cite{gans8} and its rectification is covered in Sec.~\ref{sec:spherical}. The application to the bullet cluster is covered in Sec.~\ref{sec:bullet}.
	Finally, Sec.~\ref{sec:conclusion} draws the conclusions.

\section{The MBG theory}
\label{sec:MBG}
\subsection{An overview of the MBG theory}
\label{sec:overview}
A brief overview of the MBG formulation developed in Ref.~\cite{gans8}, is now presented.
We use the letters, $a$ and $b$, as indices for the 5-D space-time, i.e., $a$, $b$ = 0, 1, 2, 3, 4, with the index 0 referring to the temperature dimension, and the index 1 referring to the time dimension,
The letters, $\mu$ and $\nu$, are indices for the 4-D Lorentzian space-time, i.e., $\mu$, $\nu$ = 1, 2, 3, 4.
A superscript, $(N)$, within brackets, refers to $N$ dimensional space. For example, $\nabla^{(4)}_{\mu}$, refers to the covariant derivative in 4-D space-time.
	Based on the theory developed in Ref.~\cite{gans7}, we consider the 5-D metric, $g^{(5)}_{ab}$, as:
            \begin{equation}
		    \label{eq:5Dmetric}
		    g^{(5)}_{ab} = 
                \left[ \begin{array}{c c}
			s^2(\bx,t)& 0\\
				    0 & g^{(4)}_{\mu\nu}\\
                \end{array} \right ],
            \end{equation}
where, $g^{(4)}_{\mu\nu}$, is the usual metric tensor in 4-D space-time, and is purely due to gravitational fields, while, $s(\bx,t)$ captures the variation in the inverse temperature, $\beta$.
The sign convention used is (+,-,+,+,+).
	For a system in thermal equilibrium, the Einstein field equation in 5-D space-time-temperature is given by~\citep{gans7}:  
\begin{equation}
	\label{eq:EFE5D}
	R^{(5)}_{ab} - \frac{1}{2}g^{(5)}_{ab} R^{(5)} = \frac{8\pi G}{c^4} T^{1(5)}_{ab},
\end{equation}
where, the stress energy tensor, $T^{1(5)}_{ab}$, is given by:
\begin{equation}
	T^{1(5)}_{ab} = (\rho + \frac{P_1}{c^2})u_a u_b + P_1g^{(5)}_{ab},
\end{equation}
	with, $\rho$ being the density, and $P_1$, the pressure.
The Ricci tensor, $R^{(5)}_{ab}$, can be expressed in terms of the 4-D covariant derivative operator, $\nabla^{(4)}_{\mu}$, and the 4-D Ricci tensor, $R^{(4)}_{\mu\nu}$, as:
            \begin{equation}
		    \label{eq:ricci5D}
		    R^{(5)}_{ab} = 
                \left[ \begin{array}{c c}
			R_{\beta\beta} & 0\\
				    0 & R^{(4)}_{\mu\nu} - \frac{1}{s}\nabla^{(4)}_{\mu} \nabla^{(4)}_{\nu} s\\
                \end{array} \right ],
            \end{equation}
where, 
	$ R_{\beta\beta} = -s\nabla^{(4)\mu} \nabla^{(4)}_{\mu} s $.
	In contrast, for systems that are non-interacting and in complete non-equilibrium, the particles behave as if there are no other particles. 
Without an ensemble or temperature concept, the temperature dimension becomes meaningless. 
To model this scenario, let us take the thermal gradient in the zero limit, i.e., $\partial_{\mu}s \rightarrow 0$, followed by $s= 0$. 
A system with zero temperature cannot have thermal gradients. This mandates $\partial_{\mu}s \rightarrow 0$. 
Subsequently, assigning $s=0$, eliminates the temperature dimension from the metric $dS^2 = s^2d\beta^2 - dt^2 + dx^2 + dy^2 + dz^2$, leading to a 4-D space-time.
As, $\partial_{\mu}s \rightarrow 0$, it is evident that the 5-D Ricci tensor in Eq.~\ref{eq:ricci5D}, is reduced to a 4-D Ricci tensor. The reduction to 4-D field equations is explained in more detail in Ref.~\cite{gans7, gans8}. 
One may also refer to Ref.~\cite{gans7} for the relations between the field theories in 5-D and 4-D.
	After reduction to 4-D, the 4-D Einstein's field equations, represented in 5-D, are:
            \begin{equation}
	\label{eq:EFE4D}
                \left[ \begin{array}{c c}
			0 & 0\\
				    0 & R^{(4)}_{\mu\nu} \\
                \end{array} \right ]
		    -\frac{1}{2} \left[ \begin{array}{c c}
			0 & 0\\
				    0 & g_{\mu\nu} R^{(4)} \\
                \end{array} \right ]
   		    = \frac{8\pi G}{c^4} \left[ \begin{array}{c c}
			0 & 0\\
				    0 & T^2_{\mu\nu} \\
                \end{array} \right ].
            \end{equation}
		    The stress energy tensors, $T^1$ and $T^2$, have the same $\rho$, but different pressures, $P_1$ and $P_2$. $P_1$ is the pressure due to an ensemble interacting gravitation-ally, while $P_2 \sim 0$ in the absence of any ensemble (a single particle has no concept of pressure).

	                An equation representing the behavior of a partially thermalized system, needs to be a generalization of Eqs.~\ref{eq:EFE5D} and~\ref{eq:EFE4D}, with  Eqs.~\ref{eq:EFE5D} and~\ref{eq:EFE4D} being special cases. 
			To motivate the generalization, let us rewrite Eq.~\ref{eq:EFE5D} by taking all terms that depend on $s$ to the R.H.S.. This is motivated by the fact that $s$ is related to inverse temperature, and thus related to inverse energy, and consequently, can be interpreted to act as a source causing space-time curvature.
		    We also apply the simplification that, for a time-invariant system, $u^0 = c\frac{\partial \beta}{\partial \tau_p} = 0$, where $\beta$ is the inverse temperature, and $\tau_p$ is the proper time. Equation~\ref{eq:EFE5D} then becomes,
\begin{multline}
	\label{eq:EFE5D_2}
                \left[ \begin{array}{c c}
			0 & 0\\
				    0 & R^{(4)}_{\mu\nu} \\
                \end{array} \right ]
		    -\frac{1}{2} \left[ \begin{array}{c c}
			0 & 0\\
				    0 & g_{\mu\nu} R^{(4)} \\
                \end{array} \right ] = \frac{8\pi G}{c^4} \\
		\times \left[ \begin{array}{c c}
			\frac{s^2c^4R^{(4)}}{16\pi G} + s^2P_1 & 0\\
				    0 & T^1_{\mu\nu} + \frac{c^4\left (\frac{1}{s}\nabla_{\mu} \nabla_{\nu} s - \frac{g_{\mu\nu}}{s}\nabla^{\alpha} \nabla_{\alpha} s \right )}{8\pi G} \\
                \end{array} \right ],
\end{multline}
where, $\alpha = 1,2,3,4$, and we have skipped the superscript $(4)$ in $\nabla^{(4)}$ for simplicity of notation. 
The terms in the R.H.S. of Eq.~\ref{eq:EFE5D_2}, can be viewed as new source terms describing a thermalized system, with thermal gradients, which causes curvature of 4-D space-time.
It is now hypothesized that the source term of a partially thermalized system may be considered to be a linear combination of the source terms in Eqs.~\ref{eq:EFE4D} and ~\ref{eq:EFE5D_2}. In other words,
\begin{multline}
	\label{eq:lincomb}
	T^{p(5)}_{ab} = 
		 (1-k) \left[ \begin{array}{c c}
			0 & 0\\
				    0 & T^2_{\mu\nu} \\
                \end{array} \right ] + k \\
		     \times \left[ \begin{array}{c c}
			     \frac{s^2c^4R^{(4)}}{16\pi G} + s^2P_1 & 0\\
				    0 & T^1_{\mu\nu} + \frac{c^4 \left (\frac{1}{s}\nabla_{\mu} \nabla_{\nu} s - \frac{g_{\mu\nu}}{s}\nabla^{\alpha} \nabla_{\alpha} s \right ) }{8\pi G} \\
                \end{array} \right ],
\end{multline}
		    where, $k$ represents the degree or extent of equilibration. 
		    Additional interpretation of $k$, is provided in Sec.~\ref{sec:interpret_k}.
		    For future use, it is worth noting that one may group the pressure terms together, and define an effective pressure:
\begin{equation}
	\label{eq:Peff}
	P_{eff} = (1-k)P_2 + kP_1.
\end{equation}
The corresponding 5-D Einstein field equations for a partially thermalized system are then:
\begin{equation}
	\label{eq:EFE5D_p}
                \left[ \begin{array}{c c}
			0 & 0\\
				    0 & R^{(4)}_{\mu\nu} \\
                \end{array} \right ]
		    -\frac{1}{2} \left[ \begin{array}{c c}
			0 & 0\\
				    0 & g_{\mu\nu} R^{(4)} \\
		    \end{array} \right ] 
		    = \frac{8\pi G}{c^4} T^{p(5)}_{ab}.
\end{equation}

For very small $k$ (as in a galaxy, or the bullet cluster with hindsight), 
one can neglect factors like $kP_1$, and given that $P_2 \sim 0$,
Eq.~\ref{eq:EFE5D_p} can be solved to obtain~\cite{gans8}:,
\begin{equation}
	\label{eq:gal3}
	R_{\mu\nu} - \frac{k}{s}\nabla_{\mu}\nabla_{\nu}s - \frac{1}{2}g_{\mu\nu}R^{(4)} + \frac{kg_{\mu\nu}}{s}\nabla^{\alpha}\nabla_{\alpha}s 
	= \frac{8\pi G}{c^4} \rho u_{\mu} u_{\nu}.
\end{equation}

If the Lorentz dilation factor $=\gamma \approx 1$, $P_2 \approx 0$ (for a non-thermalized single particle) and in the weak field limit, the following relation was derived in Ref.~\cite{gans8}:  
\begin{equation}
	\label{eq:gal8}
	\nabla^2 \phi(\bx) =   4\pi G \rho(\bx) + kc^2\frac{1}{2s(\bx)}\nabla^2s(\bx), 
\end{equation}
where, $\frac{2\phi}{c^2} = g^{(5)}_{11} - \eta^{(5)}_{11}$, and $\eta^{(5)}$ is the metric for a flat 5-D space-time-temperature.
\subsection{Spherical symmetric case}
\label{sec:spherical}

We now proceed with a simpler approach to solving the field equations compared to Ref.~\cite{gans8}.
Contracting the indexes in Eq.~\ref{eq:gal3}, one obtains:
\begin{equation}
	\label{eq:galR}
	R = \frac{8\pi G}{c^2}\rho + \frac{3k}{s}\nabla^{\alpha}\nabla_{\alpha} s.
\end{equation}
Substituting $R$ from Eq.~\ref{eq:galR} in Eq.~\ref{eq:gal3}, we obtain,
\begin{equation}
	R_{\mu\nu} = \frac{8\pi G}{c^2} \rho 
	\left (\frac{u_{\mu}u_{\nu}}{c^2} + \frac{g_{\mu\nu}}{2} \right ) + 
	\frac{k}{s}\nabla_{\mu}\nabla_{\nu}s + 
	\frac{k}{2s}g_{\mu\nu}\nabla^{\alpha}\nabla_{\alpha} s.
\end{equation}
For a spherically symmetric, weak field case, we consider the metric:
\begin{multline}
	\label{eq:sphericalmetric}
	dS^2 = (-1 + h_{11})c^2 dt^2 + (1 + h_{22})dr^2 \\
	+ r^2(1 + h_{33}) \left ( d\theta^2 + sin^2\theta d\psi^2 \right ).
\end{multline}
To terms linear in $h_{\mu\nu}$, we obtain the set of equations:
\begin{eqnarray}
	\label{eq:Rval_1}
	R_{11} = \frac{4\pi G}{c^2} \rho - \frac{k}{2s} \nabla^2 s,\\
	\label{eq:Rval_2}
	R_{22} = \frac{4\pi G}{c^2} \rho + \frac{k}{s}\frac{\partial^2s}{\partial r^2} + \frac{k}{2s} \nabla^2 s,\\
	\label{eq:Rval_3}
	R_{33} = \left [ \frac{4\pi G}{c^2}\rho + \frac{k}{2s}\nabla^2s \right ] r^2,\\
	\label{eq:Rval_4}
	R_{44} = \left [ \frac{4\pi G}{c^2}\rho + \frac{k}{2s}\nabla^2s \right ] r^2 sin^2\theta.
\end{eqnarray}
In the weak field limit, we need consider the Ricci tensor to only the first order terms in $h_{\mu\nu}$.
Then, the components of the Ricci tensor are:
\begin{equation}
	\label{eq:R11_expression}
	R_{11} = -\frac{1}{2}\nabla^2 h_{11},
\end{equation}
\begin{equation}
	\label{eq:R22_expression}
	R_{22} = -\nabla^2 h_{33} + \frac{1}{r} \frac{\partial h_{22}}{\partial r} + \frac{1}{2}\frac{\partial^2 h_{11}}{\partial r^2},
\end{equation}
\begin{multline}
	\label{eq:R33_expression}
	R_{33} = \\ r^2 \left (-\frac{1}{2}\nabla^2 h_{33} - \frac{1}{r} \frac{\partial h_{33}}{\partial r} - \frac{h_{33}}{r^2} + \frac{1}{2r} \frac{\partial h_{22}}{\partial r} + \frac{1}{2r}\frac{\partial h_{11}}{\partial r} \right ),
\end{multline}
\begin{equation}
	\label{eq:R44_expression}
	R_{44} = R_{33} sin^2\theta.
\end{equation}
From Eq.~\ref{eq:Rval_1} and ~\ref{eq:R11_expression},  
\begin{equation}
	\label{eq:R11b}
	-\frac{1}{2} \nabla^2 h_{11} = \frac{4\pi G}{c^2} \rho - \frac{k}{2s} \nabla^2 s.
\end{equation}
Substituting $h_{11} = -2\frac{\phi}{c^2}$, we obtain
\begin{equation}
	\label{eq:R11_phi}
	\nabla^2 \phi = 4\pi G \rho - \frac{k c^2}{2s} \nabla^2 s.
\end{equation}
Contrasting Eq.~\ref{eq:R11_phi} with Eq.~\ref{eq:gal8}, we find that the sign of the $\frac{k c^2}{2s} \nabla^2 s$ term has changed. Thus, the simpler derivation, helps to rectify the incorrect sign of $\frac{kc^2}{2s}\nabla^2 s$ in Ref.~\cite{gans8}.
The more involved derivation in Ref.~\cite{gans8} warranted an extra constraint to be imposed. The imposed constraint was:
\begin{multline}
\label{eq:gauge}
	\nabla^2 \Delta h_s - \partial_x^2\Delta h_{22} - \partial_y^2 \Delta h_{33} - \partial_z^2 \Delta h_{44} \\
	-2\big ( \frac{\partial^2 \Delta h_{12}} {\partial x \partial y} + \frac{\partial^2 \Delta h_{23}}{\partial y \partial z} + \frac{\partial^2 \Delta h_{13}}{\partial x \partial z} \big ) = 2\Delta h_{11},
\end{multline}
which led to the incorrect sign of $\frac{kc^2}{2s}\nabla^2 s$ in Ref.~\cite{gans8}.
The current simpler derivation invalidates the need for any additional constraint.
It is however to be noted that the results of Ref.~\cite{gans8}, can be completely be reproduced albeit with a negative value of $k$. Thus the results of Ref.~\cite{gans8} related to the galaxy rotation curves, RAR and the WBS are still valid. 
The physical interpretation and the mathematical consistency of negative $k$ are explored in Sec.~\ref{sec:interpret_k}.
In fact, the negative $k$ leads to a greater physical understanding of the MBG theory.
It is further shown in Sec.~ \ref{sec:qual}, that considerations of mathematical consistency lead to a positive $k$ if gravity is repulsive.

We now proceed with the rest of the derivation.
With reference to the metric in Eq.~\ref{eq:sphericalmetric}, we combine the Eqs.~\ref{eq:Rval_2} to~\ref{eq:Rval_4},  
with the Eqs.~\ref{eq:R22_expression} to~\ref{eq:R44_expression}, to give:
\begin{equation}
	\label{eq:R22b}
	-\nabla^2 h_{33} + \frac{1}{r}\frac{\partial h_{22}}{\partial r} + \frac{1}{2}\frac{\partial^2 h_{11}}{\partial r^2} = \frac{4\pi G}{c^2} \rho + \frac{k}{s}\frac{\partial^2s}{\partial r^2} + \frac{k}{2s} \nabla^2 s,\\
\end{equation}
and
\begin{multline}
	\label{eq:R33b}
	-\frac{r^2}{2}\nabla^2h_{33} - r \frac{\partial h_{33}}{\partial r} - h_{33} + \frac{r}{2}\frac{\partial h_{22}}{\partial r} + \frac{r}{2}\frac{\partial h_{11}}{\partial r} \\
	= \left [ \frac{4\pi G}{c^2}\rho + \frac{k}{2s}\nabla^2s \right ] r^2.
\end{multline}

Equations~\ref{eq:R22b} and~\ref{eq:R33b} can be solved to obtain
\begin{equation}
	\frac{\partial (rh_{33})}{\partial r} = -r\left \{\frac{k}{s}\frac{\partial s}{\partial r} - \frac{\partial h_{11}}{\partial r} \right \},
\end{equation}
which gives,
\begin{equation}
	h_{33} =  \frac{-1}{r} \int r \left \{ \frac{k}{s} \frac{\partial s}{\partial r} - \frac{\partial h_{11}} {\partial r} \right \} dr,
\end{equation}
Subsequently,
\begin{equation}
	\nabla^2h_{33} = \frac{-1}{r} \left \{ \frac{k}{s} \frac{\partial s}{\partial r} - \frac{\partial h_{11}}{\partial r} \right \} 
	-\frac{\partial}{\partial r} \left \{ \frac{k}{s}\frac{\partial s}{\partial r} - \frac{\partial h_{11}}{\partial r} \right \}.
\end{equation}
Rearranging Eq.~\ref{eq:R22b}:
\begin{equation}
	\label{eq:R22c}
	\frac{1}{r}\frac{\partial h_{22}}{\partial r} 
	= \frac{4\pi G}{c^2} \rho + \frac{k}{s}\frac{\partial^2s}{\partial r^2} + \frac{k}{2s} \nabla^2 s + \nabla^2 h_{33} - \frac{1}{2}\frac{\partial^2 h_{11}}{\partial r^2}.
\end{equation}
From Eq.~\ref{eq:R22c}, one can easily obtain $\nabla^2 h_{22} = \frac{1}{r^2}\frac{\partial}{\partial r^2} r^2 \frac{\partial h_{22}}{\partial r}$.

We now proceed to analyze the weak gravitational lensing effect.
For a ray of light traveling along the radial direction, $d\theta^2 = d\psi^2 = 0$. Then, after substituting $dS^2=0$ in Eq.~\ref{eq:sphericalmetric},
\begin{equation}
	\label{eq:radial_speed}
	\frac{dr^2}{dt^2} \approx c^2 \left (1 - h_{11} - h_{22}\right ).
\end{equation}
Thus, the apparent or measured speed of light is affected by the sum $h_{11} + h_{22}$.
In Einstein's gravity, for a weak field, the line element can be written as:
\begin{multline}
	dS^2 = \left (-1 + h_{11}^E \right )c^2dt^2 + \left (1 + h_{rr}^E\right ) \left (dx^2 + dy^2 + dz^2 \right )\\
	 	= \left (-1 + h_{11}^E \right )c^2dt^2 + \left (1 + h_{rr}^E \right ) \left (dr^2 + r^2d\Omega^2 \right ),
\end{multline}
where $h_{rr}^E = h_{11}^E$, if the stress energy component, $T_{11} = \rho c^2$, is the only non-zero component.
Then for a ray of light traversing in the radial direction, $d\Omega^2=0$, which then gives,
\begin{equation}
	\label{eq:ein_radial_speed}
	\frac{dr^2}{dt^2} \approx c^2 \left (1 - h_{11}^E - h_{rr}^E \right ).
\end{equation}
From Eqs.~\ref{eq:radial_speed} and~\ref{eq:ein_radial_speed},
it is apparent that the effect of $h_{11} + h_{22}$ in MBG on light is similar to that of $h^E_{11} + h^E_{rr}$ in Einstein's gravity.
We also know that in Einstein's gravity,
\begin{equation}
	-\nabla^2h_{11}^E = \frac{8\pi G}{c^2} \rho.
\end{equation}
Since $h_{rr}^E = h_{11}^E$, one may write:
\begin{equation}
	\label{eq:ein_mass}
	-\nabla^2 \left (h_{11}^E + h_{rr}^E\right ) = \frac{16\pi G}{c^2} \rho.
\end{equation}
Comparing Eqs.~\ref{eq:radial_speed}, ~\ref{eq:ein_radial_speed} and~\ref{eq:ein_mass}, we can define an effective mass, $\rho_{eff}^{radial}$, such that:
\begin{equation}
	\label{eq:eff_mass_eq}
	-\nabla^2 \left (h_{11} + h_{22} \right ) = \frac{16\pi G}{c^2} \rho_{eff}^{radial},
\end{equation}
or,
\begin{equation}
	\label{eq:eff_mass_rad}
	\rho_{eff}^{radial} = -\frac{c^2}{16\pi G} \nabla^2\left (h_{11} + h_{22} \right ).
\end{equation}
On similar lines, in the angular direction, we have,
\begin{equation}
	\label{eq:eff_mass_angle}
	\rho_{eff}^{angular} = -\frac{c^2}{16\pi G} \nabla^2\left ( h_{11} + h_{33} \right ).
\end{equation}
Based on the path of light, one can define a $\rho_{eff}$ as a weighted average
\begin{equation}
	\label{eq:eff_mass}
	\rho_{eff} = \alpha\rho_{eff}^{radial} +  \beta\rho_{eff}^{angular}.
\end{equation}
Thus, in order to demonstrate that MBG can explain the weak gravitational lensing effect of the bullet cluster, it suffices to show that $\rho_{eff}$ is comparable to the mass inferred from the weak gravitational lensing.
The bullet cluster is in the plane of the sky. For a convex lensing effect, one would expect the ray of light to be more affected by $\rho_{eff}^{angular}$ than by $\rho_{eff}^{radial}$.
We shall show in Sec.~\ref{sec:bullet} that $\rho_{eff}$ is comparable to the mass determined from the weak gravitational lensing for the bullet cluster for $\alpha = 0$ and $\beta = 1$, i.e., the angular component dominates. 

Before concluding the section, it is worth mentioning that for $k=0$, one recovers the Newtonian gravity. 
	A two-body system, where there is no concept of an ensemble, illustrates a system that is quite precisely described by $k=0$.
The value of $k$ is a free parameter, and is determined by a fit to the data in this paper. 

In Ref.~\cite{gans8}, using thermodynamic criteria, it was derived that $s\propto \frac{1}{\phi}$, where $\phi$ is the gravitational potential. 
Then,
\begin{equation}
\label{eq:sphi}
	\frac{1}{s}\nabla^2 s = \phi\nabla^2\frac{1}{\phi}.
\end{equation}
In this case, Eq.~\ref{eq:R11_phi} becomes:
\begin{equation}
\label{eq:gal9}
\nabla^2 \phi +  kc^2\frac{\phi}{2}\nabla^2\frac{1}{\phi} =   4\pi G \rho.
\end{equation}
We will revisit the relation, $s \propto \frac{1}{\phi}$, in Sec.~\ref{sec:smod}.

\subsection{Interpretation of $k$ } 
\label{sec:interpret_k}

Consider a particle moving under the influence of Newtonian gravity. It would have a deterministic velocity $v(t)_{deter}$. A particle which is part of a fully thermalized system would have a random velocity expectation value $\langle v(t)_{rand} \rangle$. But a particle which has a combination of random velocity and deterministic velocity could have the final velocity as:
\begin{equation}  
	\label{eq:linvel}
	v_{final} = (1-k)v(t)_{deter} + k\langle v(t)_{rand} \rangle,
\end{equation}  
where, $k$ can be a constant and determines the degree of mixing.
However, if instead of velocities, the energies are linearly added, then the relation would be:
\begin{equation}  
	\label{eq:linenergy}
	v^2_{final} = (1-k)v(t)^2_{deter} + k\langle v(t)^2_{rand} \rangle.
\end{equation}  
In the case of the stars in a galaxy, the $k\langle v(t)_{rand}^2\rangle$ obtains a non-zero value due to the random interaction with the other stars and gasses.
Instead of adding energies which are scalars, Eq.~\ref{eq:lincomb} incorporates a linear combination of the tensor form of the energy, namely, the stress energy tensors. 

The ramifications of $v_{rand}$ and $v_{deter}$ can be different.
As an example, a bunch of particles in random thermal motion can have a tendency to fly away from each other (a negative Ricci scalar). On the other hand a bunch of particles under each other's gravitational pull, will have a radial $v_{deter}$ towards each other (a positive Ricci scalar). A real system may be a combination of both the effects. 

An explosive with an energy of explosion, E, would expand. But, the gravitational effect of energy $E$, would try to contract, although a much weaker effect. 
The resultant effects of these two phenomena is captured by the MBG theory, 

In the following subsection, we explore the mathematical and physical consistency due to the negative value of $k$

\subsubsection{Modeling of $s$}
\label{sec:smod}
We now revisit the derivation of the relation, $s \propto \frac{1}{\phi}$, in Ref.~\cite{gans8}, in the context of a negative $k$. For the sake of completeness, the arguments and derivations in Ref.~\cite{gans8} are reproduced.

For a partially thermalized system like a galaxy, the stress energy tensor is modeled as a linear combination of the stress energy tensors of a hypothetical thermalized system in local thermal equilibrium, 
and a system in complete non-equilibrium (Ref. Eq.~\ref{eq:lincomb}).
The thermalized system has a variation in inverse temperature $=s$,  
Although, $s$ appears in Eq.~\ref{eq:R11_phi}, only the properties of a galaxy can be measured. The variable, $s$, refers to the variation in the inverse temperature of a hypothetical thermalized system, and therefore cannot be measured.
Thus, it is required to relate $s$ to a physically measurable state variable of a galaxy (or a galaxy cluster or any other system), which can then be used to determine $s$, or used in place of $s$. 

Let us first consider a fully thermalized system. The energy expectation value is given by:
\begin{equation}
\label{eq:temperature}
	\langle E\rangle = \frac{1}{Z}\sum_{i}n_i E_i \exp\left (-\frac{E_i}{\sK T}\right ),
\end{equation}
where, $n_i$ particles have energy $E_i$, $Z$ is the partition function, and $\sK$ is the Boltzmann constant.
For a given distribution, $n_i$, it is a bijective mapping between $\langle E\rangle$ and temperature, $T$.
The temperature of a thermal system thus becomes a measure of the energy expectation value of the system. 
Independent of the distribution, the inverse temperature or $s$, can be generalized as a function of the variation of the inverse energy, $\langle E\rangle$. Specifically, $s = \frac{1}{f(\langle E\rangle)}$.
The function, $f(\langle E\rangle)$, itself may depend on the distribution of particles. 
In other words, different systems such as Boson gases, Fermion gases, and classical systems can have different functions.
Let $\frac{1}{\sK T(\bx)} = \beta(\bx) = s(\bx)\beta_0$, with $\beta_0$ = a constant, be the inverse temperature at a point $\bx$ in the thermalized system. One may Refer~\cite{gans6,gans7} for the relation $\beta(\bx) = s(\bx)\beta_0$.
In the case of a classical system in local thermal equilibrium, the expected energy, $\langle E_{cl}(\bx)\rangle$, of a particle at a point $\bx$, can be taken as: $\langle E_{cl}(\bx)\rangle \approx \sK T(\bx)$. 
Then, one obtains:
\begin{equation}
\label{eq:sE}
	s(\bx) \approx \frac{1}{\beta_0\langle E_{cl}(\bx) \rangle} .
\end{equation}
Thus, for a thermalized system, the energy expectation value and the temperature can be used interchangeably, as they represent the same physics.

For a partially thermalized system, we relate its pressure to the pressure of a fully thermalized system.
From Eq.~\ref{eq:Peff}, we have, $P_{eff} = (1-k)P_2 + kP_1$. 
Here, $P_1$ is the pressure of the thermalized system, with inverse temperature, $\beta(\bx) = s(\bx)\beta_0$.
Let us consider a small number of particles, $\Delta n(\bx)$, contained within a small volume, $\Delta V_{\bx}$, around a point $\bx$.
The energy expectation value, $\langle E(\bx)\rangle$, of a particle at the point $\bx$, in a partially thermalized system is then:
\begin{multline}
	\label{eq:Ea}
	\langle E(\bx) \rangle = \frac{1}{\Delta n(\bx)}\int_{\Delta V_{\bx}} P_{eff}dV\\
	= \frac{1}{\Delta n(\bx)}\int_{\Delta V_{\bx}} \left \{ (1-k)P_2 + kP_1 \right \} dV,
\end{multline}
In Sec.~\ref{sec:overview}, we had taken $P_2 \sim 0$, for a non-interacting system in complete non-equilibrium, like an isolated single particle, or a collection of isolated non-interacting particles. 
Pressure is an ensemble concept and can only be defined for an ensemble.
With $P_2 \sim 0$, we obtain,
\begin{equation}
	\label{eq:Eb}
	\langle E(\bx) \rangle = \frac{1}{\Delta n(\bx)}\int_{\Delta V_{\bx}} kP_1 dV = k\langle E_1(\bx) \rangle,
\end{equation}
where, $\langle E_1(\bx) \rangle = \frac{1}{\Delta n(\bx)}\int_{\Delta V_{\bx}} P_1 dV$, is the energy expectation of a particle within the small volume, $\Delta V_{\bx}$, existing in a thermalized system at local thermal equilibrium. 
In the case of the thermalized system being a classical system in local thermal equilibrium, we infer from Eq.~\ref{eq:sE}, $\langle E_1(\bx) \rangle \approx  \frac{1}{\beta_0 s(\bx)}$. Consequently, Eq.~\ref{eq:Eb} leads to:
\begin{equation}
	\label{eq:Ec}
	\langle E(\bx) \rangle \approx \frac{k}{\beta_0 s(\bx)}.
\end{equation}

In a gravitational system, like a galaxy or a galaxy cluster, the potential energy of a particle (star) is proportional to the gravitational potential, $\phi$, it sees due to the presence of other particles (stars or gasses) in the system. Subsequently, we take the energy expectation value, $\langle E(\bx) \rangle \propto \phi(\bx)$, where $\phi$ is the gravitational potential. 

Putting everything together, we get, $s\propto \frac{1}{\phi}$.
It is to be noted that, since $\frac{1}{s}\nabla^2s$ is scale-invariant, any constant of proportionality in $s$ is irrelevant.
In this case, Eq.~\ref{eq:R11_phi} becomes:
\begin{equation}
	\label{eq:gal10}
	\nabla^2 \phi  =   4\pi G \rho -  kc^2\frac{\phi}{2}\nabla^2\frac{1}{\phi}.
\end{equation}

Now that we have re-derived Eq.~\ref{eq:gal10}, let us examine the sign of $k$. 
The gravitational potential energy is attractive and hence, negative. Therefore, $\langle E(\bx) \rangle$ should be negative. On the other hand, the thermal energy represented by $\langle E_1(\bx) \rangle$ is positive. 
Equation~\ref{eq:Eb}, requires a negative $k$, to map a positive $\langle E_1(\bx) \rangle$ to a negative $\langle E(\bx) \rangle$.
Thus, $k$ needs to be negative for Eq.~\ref{eq:Eb} to be mathematically consistent.

Lastly, let us investigate any potential restrictions on $k$.
In Eq.~\ref{eq:Eb}, if $k \ll -1$, a finite thermal energy, $\langle E_1(\bx) \rangle$, can lead to an infinitely large gravitational energy, $\langle E(\bx) \rangle$, which is unrealistic. If we mandate $|\langle E(\bx) \rangle| < |\langle E_1(\bx) \rangle |$, then bounds on $k$ would be $-1 < k < 0$.

\subsubsection{Negative $k$ and qualitative analysis}
\label{sec:qual}
The argument in Sec.~\ref{sec:smod} demonstrates that mathematical consistency is achieved by having a negative value of $k$.
The mathematical consistency and the correct behavior of the galaxy rotation curves (Eq.~\ref{eq:R11_phi}) suggest that $k$ must be negative.

At a more intuitive level, an equilibrated system has a "grouping effect". 
The grouping effect happens due to the interaction between the thermal particles. This could be accentuated in a gravito-thermal system, where the interaction between the particles (stars or gases) is attractive.
The ensemble of stars, gas molecules, and galaxies act as a weakly bound syrup, rather than as disjointed, non-interacting individual particles.
The positive kinetic energy of particles in a partially equilibrated thermal system is expected to result in a repulsive system.
However, the random thermal velocity is a result of the tugs and pulls caused by an attractive gravitational interaction.
The dichotomy between the positive thermal energy and the attractive gravitational force is captured by a negative sign in the value of $k$.

Let us consider some examples. Consider a hyper-velocity star~\citep{hypervel} that has escaped the Milkyway galaxy. This would be an isolated star in space, not interacting with any other system in space. 
For this isolated star, $P_2 \sim 0$ and $k \sim 0$.   
Even while the hyper-velocity star is still within the galaxy, the ratio $\frac{v(t)^2_{deter}}{\langle v^2(t)_{rand} \rangle}$ is much higher, compared to other stars. Hence one may expect $|k|$ to be very small, and gradually decrease to zero as the star moves out of the galaxy.
For the other normal stars in the galaxy, there would be a non-zero galactic pressure $\approx kP_1$, where $P_1$ is the pressure of a hypothetical system with the stars in thermal equilibrium. 
The pressure, $kP_1$, is the result of the tugs and pulls that each star experiences inside the galaxy. At the galactic core, one would expect the gravitational potential, $\phi$, to be the highest. 
At that point, the pulls and tugs would also be high, resulting in high galactic pressure.
The higher pressure can be attributed to either a larger value of $k$ or a larger value of $P_1$.
In the current formalism, $k$ is kept constant and $P_1(\bx)$ is altered.
To sum up, the galactic system is considered to be the superposition of two systems.
\begin{enumerate}
	\item A non-equilibrium system with $P_2 \sim 0$. 
		Each star in the galaxy would be viewed as a single isolated particle. For each star, $v^2_{deter} > 0$ and $v^2_{rand}=0$. 
In a magical way, there are no tugs and pulls between the stars!
An analogy would be a set of balls tied to a pole by strings, and rotating around it. 
Since there are no strings between the balls, there is no interaction between them.
	\item An hypothetical equilibrated system with pressure, $P_1(\bx)$, and temperature, $T(\bx)$. The pressure, $P_1(\bx)$, and temperature, $T(\bx)$, are higher near the interior.
\end{enumerate}
In regions of an equilibrated system with higher pressure, the temperature is expected to increase.
Therefore, the inverse temperature and consequently, $s(\bx)$, would be reduced in those areas.
It should be noted that the temperature, $T(\bx) = \frac{1}{s(\bx)\beta_0}$, is not the temperature one would observe if one were to place a thermometer in the galaxy. 
The temperature of a galaxy in the context of the MBG formalism, is not the same as that of a star or the interstellar space.
Eq.~\ref{eq:temperature} determines the temperature, $T$, of the hypothetical equilibrated system, where, the distribution $n_i$, represents the distribution of stars.
The temperature of the galaxy is then $kT(\bx)$, and $s(\bx) = \frac{1}{\beta_0 T(\bx)}$.
The explanation given above has been captured in a more formal way in Sec.~\ref{sec:smod}. The derivation in Sec.~\ref{sec:smod}, shows that $\phi \propto T$.

In the case of a galaxy cluster like the bullet cluster, the ensemble is not the collection of stars, but a collection of galaxies and gasses. The gasses and the galaxies interact with each other and are responsible for the tugs and pulls between each other, leading to a weakly bound "syrup".
In fact, one may identify two distinct sub-systems
\begin{enumerate}
\item The galaxies rotating around each other forming a cluster. This is again a small ensemble, with each galaxy pulling and tugging each other. 
\item The ICG gasses, which fill the space around the galaxy cluster. 
\end{enumerate}
It seems likely, that the ICG would form a much more equilibrated system than the galaxy cluster itself. Therefore, the value of $k$ for the ICG should be much higher.
The galaxy cluster may be categorized as a partial thermalized system, but it may be closer to a non-equilibrated system.
	Referring to Eq.~\ref{eq:linenergy}, the $k\langle v(t)^2_{rand}\rangle$ would be significantly high for the gas molecules in the ICG, which implies a higher value of $k$.
	Since the ICG can be considered static, $v(t)^2_{deter}$ would be zero. Instead, we may consider the equivalent gravitational potential energy for the deterministic component. 
At present, the current formalism lacks the generalization to handle multiple $k$s.
Hence, as an approximation, in this work we treat both the galaxy cluster and the ICG on the same footing with the same value of $k$.
The gravitational force exerted on the ICG gasses near the galaxy clusters would be stronger.
The density of the gas molecules in this region should increase, leading to an increased inter-molecular interaction and random thermal motion of the gas molecules.
As a result, the ICG near the brightest cluster galaxies (BCG) experiences a higher pressure, $kP_1$, and temperature, $kT(\bx)$.
Again, the gravitational interaction between gas molecules results in a negative gravitational potential, $\phi$, and consequently, the energy $\langle E(\bx) \rangle$ is negative.
Based on Eq.~\ref{eq:Eb}, $k$ would need to be negative to map  a positive
$\langle E_1(\bx) \rangle $, to a negative $\langle E(\bx) \rangle$. 

Finally, let us examine two extreme cases of 
\begin{itemize}
\item very strong gravity and, 
\item repulsive gravity.
\end{itemize}
In a galaxy, let us boost $G$, the Newton's gravitational constant, and consider the stars as hard spheres. 
For a sufficiently large $G$, the gravitational forces of attraction between the stars become significantly larger, and the stars would abut each other. 
The galaxy could start to resemble a rigid body more and more.
The velocity of stars $=v \propto r$, where $r$ is the radial distance.
A very large negative value of $k$ could lead to a solution that approximates $v\propto r$.
	Let us now visit the second case, and make gravity repulsive by changing the sign of $G$. The gravitational potential, $\phi$, becomes positive. Consequently, the gravitational energy, $E(\bx)$, becomes positive.
Then, in Eq.~\ref{eq:Eb}, a positive $k$ is required to map a positive $\langle E_1(\bx) \rangle$ to a positive $\langle E(\bx) \rangle$.
A positive $k$ reverses the effect of $\frac{kc^2}{2}\phi \nabla^2 \frac{1}{\phi}$, and the galaxy explodes.
\section{Bullet Cluster: A proof of concept}
\label{sec:bullet}
\subsection{Modeling of the system}
The MBG formalism is now applied to the bullet cluster.
The terms, ICG, plasma, and gas are used interchangeably in this work.
The velocity of the bullet cluster after the collision = $v_{bullet} \approx 4700 km/s$~\citep{bullet4}. Since $v_{bullet} \ll c$, static conditions can be applied, since, all the time derivatives are close to zero. 
Thus, Eq.~\ref{eq:gal9} is applicable to the bullet cluster.

 The crux of the solution lies in the fact that the MBG theory, not only depends on the baryonic distribution, but also on the second derivative of the gravitational potential, $\phi$.
 Galaxies would tend to attract the ICG towards it, due to the gravitational pull. 
 As a consequence, the density of the gas can be expected to be higher near the galaxies, and decrease radially away from them. 
 It is shown, that variations in gas densities, lead to variations in the gravitational potential, $\phi$, resulting in the manifestation of the term, $\frac{kc^2}{8\pi G}\phi \nabla^2 \frac{1}{\phi}$.
The galaxy clusters themselves would be confined to a finite region leading to additional contribution to the term $\frac{kc^2}{8\pi G}\phi \nabla^2 \frac{1}{\phi}$.
The galaxy rotation curves were mainly affect by $h_{11}$ in Eq.~\ref{eq:sphericalmetric}.
The perturbation, $h_{11}$ depends exclusively on $\rho$ and  $\frac{kc^2}{8\pi G}\phi \nabla^2 \frac{1}{\phi}$.
However, as seen in Eq.~\ref{eq:radial_speed}, the path of light depends on both $h_{11}$ and $h_{22}$ (or $h_{33}$). The effective mass, $\rho_{eff}$, captures this effect, as seen in Eqs.~\ref{eq:eff_mass_rad} and~\ref{eq:eff_mass_angle}.
We now show that the effective mass in Eq.~\ref{eq:eff_mass}, $\rho_{eff}$, is comparable to the mass inferred via the weak gravitational lensing.

The 3-D distribution of the ICG is necessary for determining the gas density.
 Unfortunately, only the surface density of the ICG can be observed. The 3-D distribution of the ICG needs to be calculated. 
 To illustrate the concept, we employ a phenomenological model of the 3-D gas distribution, which is derived in Sec.~\ref{sec:rho}. 
As a first approximation, we use the following ICG distribution, with the function $f(r)$ in Eq.~\ref{eq:Brho2} approximated as a constant:
 \begin{equation}
 \label{eq:rhogas}
	 \rho(x,y,z) = \sum_{i=1}^{N} A_i(x,y) \rho_{0i}(|\vec{\br} - \vec{\br}_i|),
 \end{equation}
 where, 
 \begin{itemize}
	 \item $\rho_{0i}(\br - \br_i) = \exp\left \{-c_i(|\vec{\br} - \vec{\br}_i)|\right \}$ (note: this contains the approximation $f(r) = c_i$), 
	 \item $|\vec{\br} - \vec{\br}_i| = \sqrt{(x-x_i)^2 + (y-y_i)^2 + (z- z_i)^2}$,
	 \item $(x_i, y_i, z_i)$ = location of $i^{th}$ galaxy. and,
	 \item $c_i$ are constants (= 0.4 based on the fit to data), 
	 \item $N$ = number of galaxy clusters, and,
	 \item $A_i(x,y)$ is a scaling factor to be determined.
 \end{itemize}
 Let $\Sigma(x,y)$ be the surface density of the ICG. Substitute, $A_i(x,y) = B(x,y)\left (\frac{M_i}{\sum_i M_i} \right )$, where $M_i$ is the mass of the $i^{th}$ galaxy,
 \begin{equation}
	 \label{eq:scale}
	 B(x,y) =  \frac{\Sigma(x,y)}{\int \sum_i \left [ \left ( \frac{M_i}{\sum_i M_i} \right )\rho_{0i}(x,y,z)\right ] dz}.
 \end{equation}
The normalization in Eq.~\ref{eq:scale} ensures that $\int \rho(x,y,z) dz = \Sigma(x,y)$.
The scale factor, $B(x,y)$, may be interpreted as the modulation of the intrinsic ICG distribution, $\rho_{0i}$, due to the galactic collision.

The position of the BCG and the plasma gas peaks, given in table~\ref{tab:coord}, were obtained from Ref.~\cite{bullet4}.
\begin{center}
\begin{table}
\begin{tabular}{ |c|c|c|  }
\hline
		Component  & RA & Dec \\
		\hline
		Main BCG & 06:58:35.3 & -55:56:56.3  \\
		Main cluster plasma & 06:58:30.2 & -55:56:35.9 \\
		Sub cluster plasma & 06:58:21.2 & -55:56:30.0 \\
		Sub cluster BCG & 06:58:16.0 & -55:56:35.1 \\

\hline
\end{tabular}
	\caption{Table depicting the RA and Dec coordinates of main components used in the modeling~\citep{bullet4}. }
\label{tab:coord}
\end{table}
\end{center}

Since, the main and the sub cluster galaxies are lumped as a) the main cluster BCG and b) the sub cluster BCG, we take $N$ = number of galaxy clusters = 2.
The mass of the main cluster BCG was taken as $M_{BCG1} = 5.5\times 10^{12} M_{\odot}$, while the sub cluster BCG mass was taken as $M_{BCG2} = 2.7\times 10^{12} M_{\odot}$~\citep{bullet4}. 
In terms of distribution, the main and sub cluster BCG were assumed to be spherically symmetric Gaussian with standard deviations of 144kpc and 115kpc respectively. The main objective of this distribution is to have a smooth curve and confine most of the BCG mass within a finite volume.
The mass of the ICG (plasma gas) was taken as $3.87\times10^{14} M_{\odot}$~\citep{moffat}.
The sampling rate = 13.6kpc, and value of $k = -3.5609\times 10^{-3}$.
The distributions of $\Sigma(x,y)$ and $\kappa$ were obtained from "https://flamingos.astro.ufl.edu/1e0657/data/". The $\kappa$ distribution indicates the mass distribution obtained from weak gravitational lensing~\citep{bullet4}.
The $\kappa$ peaks (almost) coincide with the main and sub BCG peaks, and are different from the ICG peaks.
The sampling rates and orientation of $\kappa$ and $\Sigma(x,y)$ were different in the database.
Consequently, the $\kappa$ values were interpolated and rotated to align with the $\Sigma(x,y)$ distribution, with the help of table~\ref{tab:coord}.

The bullet cluster as a whole is not spherically symmetric. However, the formal development in Sec.~\ref{sec:spherical} is for a spherically symmetric system. The term, $\frac{1}{s}\nabla_{\mu}\nabla_{\nu} s$, is non-diagonal for a generic case, and the field equations become formidable. 
Fortunately, the spherically symmetric equations can still be used to approximately model the effects of the gas distribution close to the two BCGs. 
The other issue is the non-linearity of the term, $\phi\nabla_{\mu}\nabla_{\nu} \frac{1}{\phi}$ (= $\frac{1}{s}\nabla_{\mu}\nabla_{\nu} s$)  The non-linearity implies, that it would not be possible to solve the effect of the gas distribution near one BCG, and linearly add the potential, $\phi_i$, due to each BCG. 
However, if $\phi_2 \ll \phi_1$,  
\begin{equation}
	\label{eq:linapprox}
	(\phi_1 + \phi_2)\nabla_{\mu}\nabla_{\nu} \frac{1}{\phi_1 + \phi_2} 
\approx	\phi_1 \nabla_{\mu}\nabla_{\nu} \frac{1}{\phi_1 }.
\end{equation}
If the two BCGs are sufficiently far away, then $\phi$ due to the gas distribution around one BCG, would be very small near the other BCG. 
Hence, we may use the approximation in Eq.~\ref{eq:linapprox}.
This approximation would be unlikely to  affect the accuracy of the results near the BCGs. However, one should expect a lack of accuracy in the middle region between the two BCGs. 
 Based on the approximation in Eq.~\ref{eq:linapprox}, the potential and the effective mass due to each BCG is calculated standalone. Subsequently, they are combined as:
            \begin{equation}
		    \label{eq:combineBCG}
		    \rho_{eff}(\br) = 
		                    \begin{cases}
					    \rho_{eff}^{BCG1}(\br); & \text{if~} | \br - \br_{BCG1}| < \frac{M_{BCG1}}{M_{BCG1} + M_{BCG2}} \Delta \br,\\
					     & \\
					    \rho_{eff}^{BCG2}(\br); & \text{if~} | \br - \br_{BCG2}| < \frac{M_{BCG2}}{M_{BCG1} + M_{BCG2}} \Delta \br,\\
		\end{cases}  
            \end{equation}
	    where, $\Delta \br= \br_{BCG2} - \br_{BCG1}$.
A smoothing is also performed at the meeting point of the two regions in Eq.~\ref{eq:combineBCG}, in order to smooth out any artificial abrupt changes.

\subsection{Simulation results }
\begin{figure}
\includegraphics[width = 80mm,height = 80mm]{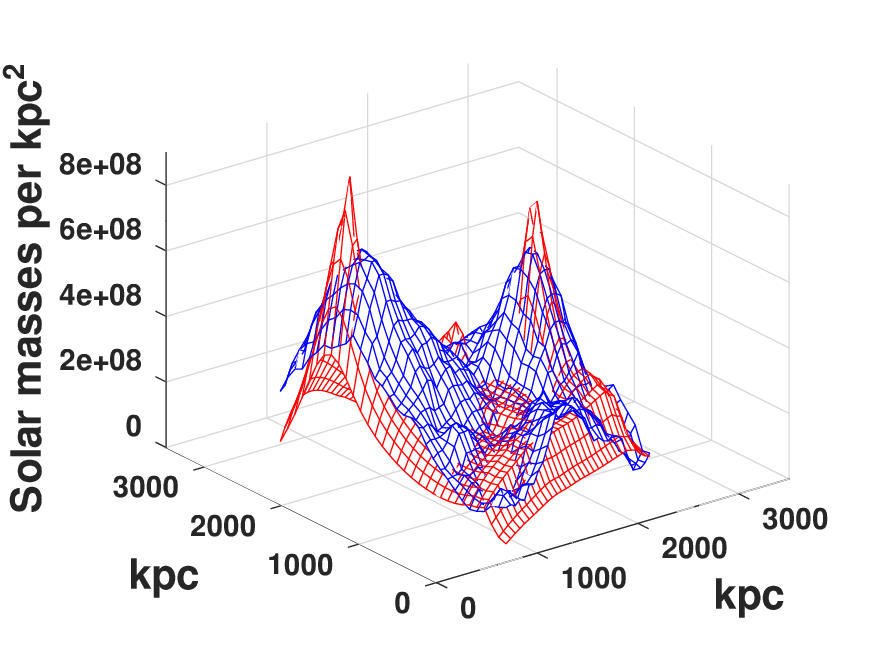}
	\caption{Bullet cluster: Superimposition of the surface mass due to the effective mass, $\rho_{eff}^{angular}$ from MBG (red) upon the $\kappa$ map from weak gravitational lensing (blue)~\citep{bullet4}.
	} 
\label{fig:bullet_3D}
\end{figure}

\begin{figure}
\includegraphics[width = 80mm,height = 80mm]{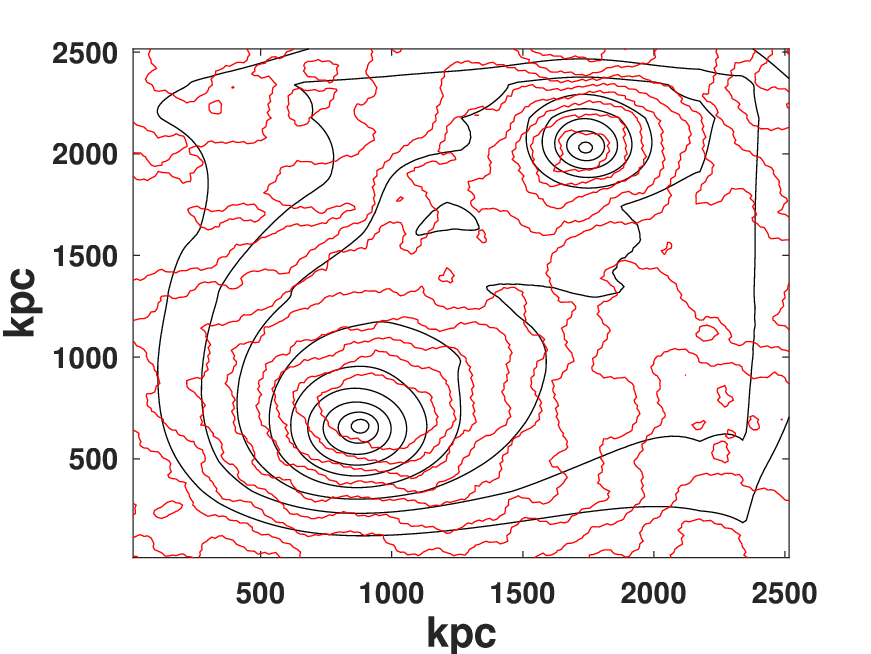}
	\caption{Bullet cluster: Contours of the surface mass due to the effective mass, $\rho_{eff}^{angular}$, from MBG (black) and the contours of the $\kappa$ map from weak gravitational lensing (red)~\citep{bullet4}.
	} 
\label{fig:bullet_contour}
\end{figure}

\begin{figure}
\includegraphics[width = 80mm,height = 80mm]{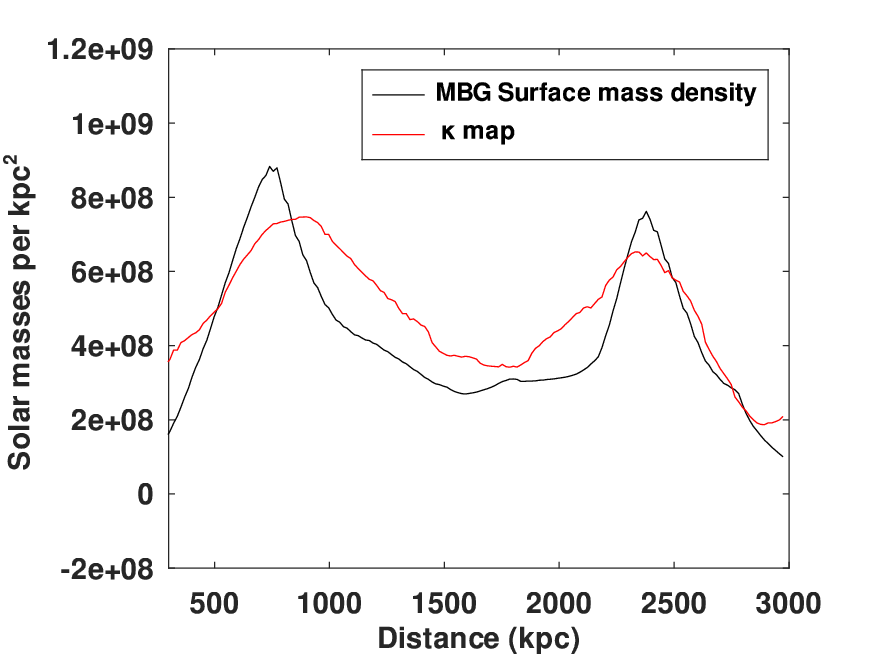}
	\caption{Bullet cluster: Cross sectional view of Fig.~\ref{fig:bullet_3D}. Comparison between the surface mass due to the effective mass, $\rho_{eff}^{angular}$, from MBG (black) and the $\kappa$ map from weak gravitational lensing (red)~\citep{bullet4}.
	} 
\label{fig:bullet_line}
\end{figure}

We now examine the simulation results.
The numerical simulation started with a grid size 360x360x360 and a procedure similar to Ref.~\cite{gans8} was followed.
The effective mass, $\rho_{eff}$, is calculated based on the discussions in Sec.~\ref{sec:spherical}. 
The surface distribution is determined by integrating the effective mass along the $z-axis$.
Figure~\ref{fig:bullet_3D} shows the surface distribution of the effective mass, $\rho_{eff}^{angular}$, obtained from the MBG theory, overlaying the $\kappa$ distribution.
Figure~\ref{fig:bullet_contour}, provides a contour view of the effective mass, $\rho_{eff}^{angular}$, and the $\kappa$ distribution. The red curves depict the $\kappa$ variation, while the black curves depict the effective mass variation. They are seen to be similar.
Finally, Fig.~\ref{fig:bullet_line} depicts a cross-sectional view of the effective mass, $\rho_{eff}^{angular}$, and the $\kappa$ distribution. 
The peaks of the effective mass more or less coincide with the peaks of $\kappa$.
The MBG theory undeniably and unequivocally shows the peaks of the weak gravitational lensing effect to be near the BCGs.
Thus, while the peak of the baryonic mass is in the middle, the weak gravitational lensing is concentrated near the BCGs, due to the gradients in the ICG distribution near the BCGs, and partially due to the location and distribution of BCGs themselves.


\begin{figure}
\includegraphics[width = 80mm,height = 80mm]{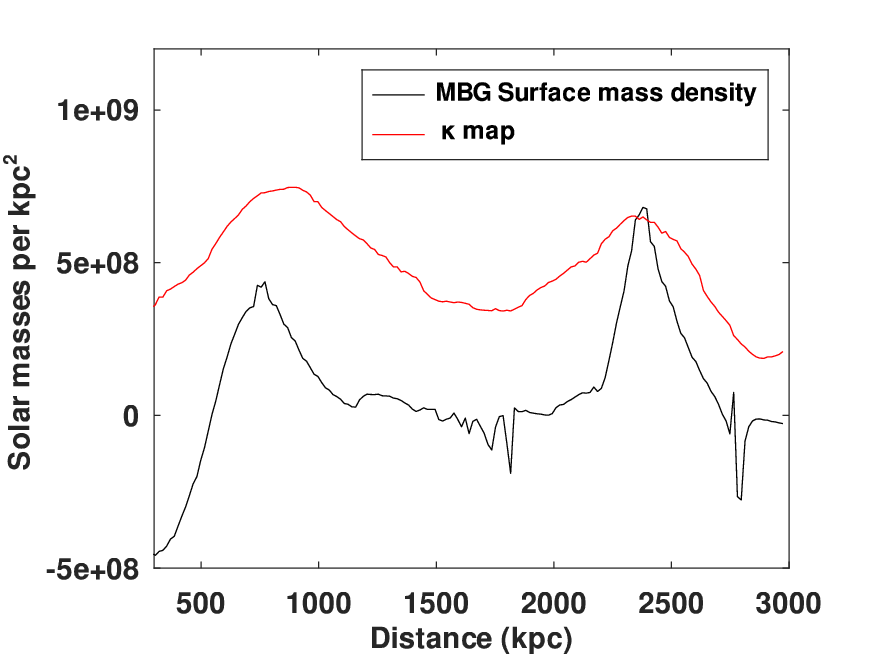}
	\caption{Radial effective mass due to Bullet cluster: Cross sectional view of the effective mass in the radial direction. Comparison between the surface mass due to the effective mass, $\rho_{eff}^{radial}$, from MBG (black) in the radial direction and the $\kappa$ map from weak gravitational lensing (red)~\citep{bullet4}.
	} 
\label{fig:bullet_line_radial}
\end{figure}

	For the sake of completeness, the cross sectional view of the effective mass in the radial direction, $\rho_{eff}^{radial}$, is depicted in Fig.~\ref{fig:bullet_line_radial}.

	Before concluding, it is worth mentioning that in the context of Eq.~\ref{eq:linenergy}, the gravitational potential energy of the ICG, near the BCG, would replace the deterministic kinetic energy term, $v(t)^2_{deter}$. Also, $v^2_{final}$ would be replaced by total energy. 

	\section{Conclusion}
	\label{sec:conclusion}
	The MBG theory, with its genesis in modeling thermal systems at the quantum mechanical level, is able to reproduce the weak gravitational lensing effect of the bullet cluster to a reasonable extent, without the need for yet undiscovered baryonic matter or dark matter.
The effective mass distribution depends on the second derivative of the gravitational potential, $\phi$, and can differ considerably from the baryonic mass distribution. This can significantly impact the weak gravitational lensing effect.

The MBG theory was previously demonstrated to reproduce the galaxy rotation curves, the RAR relation, and be in accordance with the WBS observations.
	Thus, with the reproduction of the bullet cluster weak lensing effect, the MBG theory is able to reproduce several astrophysical phenomena.

While, the thermal geometric effects have been investigated in this work, more investigation on the thermal geometric effects is required to understand all its properties.


\bibliographystyle{elsarticle-num}
\bibliography{bullet}

\appendix
\input{get_rho}


\end{document}

%% file: get_rho.tex
\section{The gas density}
\label{sec:rho}
\begin{figure}
\includegraphics[width = 60mm,height = 80mm]{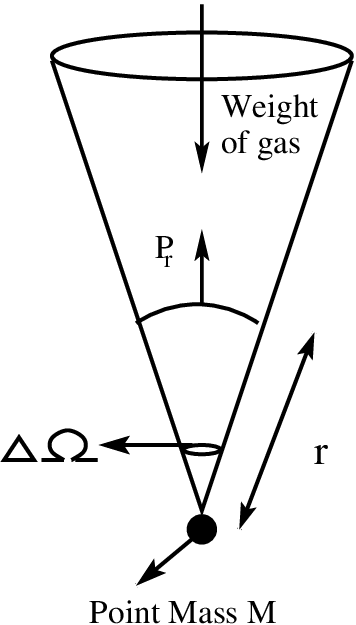}
	\caption{The pressure at $r$ should support the column of gas above it, contained in the solid angle $d\Omega$}.
\label{fig:rho}
\end{figure}

We now determine the density of gas as a function of distance, r, from a point mass $M$.  We assume spherical symmetry.
Referring to Fig.~\ref{fig:rho}, we wish to determine the density $\rho(r)$ at a distance $r$. Let pressure at a radial distance, $r$, be $P_r$. This pressure should support the mass of the gas in the conical volume of solid angle $\Delta \Omega$ above it.
Then,
\begin{equation}
\label{eq:press_rho}
	P_r r^2\Delta \Omega = \int_r^{\infty} \rho(\zeta) g(\zeta) \zeta^2 d\zeta \Delta \Omega,
\end{equation}
where, $g(r)$ is the acceleration due to gravity at a distance, $r$.
Taking $P_r = \rho(r)RT(r)$, where, $R$ = universal gas constant and, $T$ = temperature, we have,
\begin{equation}
	\rho(r)RT(r) r^2 = \int_r^{\infty} \rho(\zeta) g(\zeta) \zeta^2 d\zeta.
\end{equation}
Differentiating both sides, w.r.t. $r$,
\begin{multline}
	\frac{d \rho(r)}{dr} R T(r) r^2 + 2\rho R T(r) r +  \frac{d T(r)}{dr} R \rho(r) r^2\\
	= -\rho(r)g(r)r^2.
\end{multline}
Rearranging and solving we get
\begin{equation}
\label{eq:Arho}
	\rho(r) = A \exp \left \{-\int \left ( \frac{g(r)}{RT(r)} + \frac{1}{T(r)}\frac{dT(r)}{dr} + \frac{2}{r} \right )dr \right \}, 
\end{equation}
where, $A$ is a constant of integration.
The gravitational force, $F\,\propto\, M$, where $M$ is the galactic (point) mass. Again $F = \frac{d P_r}{dr}$. 
But from Eq.~\ref{eq:press_rho} and ~\ref{eq:Arho}, $P_r \propto A$. All this implies $M \propto A$.
Thus, in Eq.~\ref{eq:Arho}, we replace $A\rightarrow B M$, where $B$ is a constant. Then,
\begin{equation}
\label{eq:Brho2}
 \rho(r) = BM\exp \left \{-\int f(r) dr \right \}, 
\end{equation}
	where, $f(r) =  \frac{g(r)}{RT(r)} + \frac{1}{T(r)}\frac{dT(r)}{dr} + \frac{2}{r}$. 
	Both $g(r)$ and $T(r)$ are expected to decrease with $r$, and hence, $\frac{g(r)}{T(r)}$ may have a gentler slope than $g(r)$.
The acceleration due to gravity, $g(r)$, itself would now be based on the MBG theory, and not the Newtonian gravity.
Eventually, the highest power of $r$ in $f(r)$, will dominate the behavior of $f(r)$ at large $r$. In fact, it was seen in Sec.~\ref{sec:bullet}, that $f(r)$ = constant reproduces the observed data to a reasonable extent.